\documentclass[notitlepage,preprint]{article}
\usepackage[utf8]{inputenc} 
\usepackage[bitstream-charter]{mathdesign} 
\usepackage[colorlinks=true,urlcolor=blue,linkcolor=blue, citecolor=blue]{hyperref}
\usepackage{graphicx}
\usepackage[a4paper,left=3.5cm, right=3.5cm, top=3cm, bottom=3cm]{geometry}
\usepackage{authblk}

\usepackage{amsthm}
\usepackage{amsmath}

\title{Cosmic Rays Induced Background Radiation on Board of Commercial Flights}

\author[1]{Sergio Pinilla}
\author[2,1]{Hernan Asorey\thanks{Corresponding author: \href{mailto://hasorey@uis.edu.co}{hasorey@uis.edu.co}}}
\author[1]{Luis A. N\'u\~nez}

\affil[1]{Escuela de F\'isica, Universidad Industrial de Santander,
Bucaramanga, Colombia}
\affil[2]{Laboratorio de Detecci\'on de Part\'iculas y Radiaci\'on, Centro At\'omico Bariloche \& Instituto Balseiro, San Carlos and Bariloche, Argentina}

\date{\today}
\begin{document}
\maketitle

\begin{abstract}
	The aim of this work is to determine the total integrated flux of cosmic
radiation which a commercial aircraft is exposed to along specific flight
trajectories. To study the radiation background during a flight and its
modulation by effects such as altitude, latitude, exposure time and transient
magnetospheric events, we perform simulations based on Magnetocosmics
and CORSIKA codes, the former designed to calculate the geomagnetic
effects on cosmic rays propagation and the latter allows us to simulate the
development of extended air showers in the atmosphere. In this first work, by
considering the total flux of cosmic rays from $5$\,GeV to $1$\,PeV, we
obtained the expected integrated flux of secondary particles on board of a
commercial airplane during the Bogot\'a-Buenos Aires trip by point-to-point
numerical integration.
\end{abstract}

\section{Introduction}\label{sec:introduction}

A nearly constant flux of particles of Solar, Galactic or Extragalactic origin
arrives to the near-Earth environment. These particles, known as \textit{cosmic
rays} (CR), reach our planet with energies that vary over a wide range,
beginning at $10^5$\,eV for solar wind particles and ending beyond
$10^{20}$\,eV for extragalactic cosmic rays. The lower energy CR are deflected
by the Earth magnetosphere, but as the particles energy becomes higher, they
can go trough the magnetosphere and thus, they are able to penetrate into the
atmosphere. When one of this \textit{primaries} collides with a constituent
nucleus of the air (typically a nitrogen atom), a cascade of \textit{secondary
particles}, called an \textit{Extensive Air Shower} (EAS), is generated. 

As the altitude increases, the atmosphere protective layer becomes thinner and
less dense, and this is the main reason to consider the incidence of cosmic
radiation over a commercial aircraft flying between $10$\,km to $12$\,km, since
at those levels the background radiation produced by secondary particles is
much higher than at ground level. Besides altitude, there are other factors
that may affect the dose received by the airplane electronics, the passengers
and the aircraft crew, such as the geomagnetic coordinates of the plane
trajectory, different space weather conditions, and of course the exposure.

This phenomenon has been investigated only in the last two decades and it has
become an occupational health issue in some countries (see, for example
\cite{bottollier-depois_comparison_2012}). To calculate the number of particles
incident over an aircraft flying along different routes, simulations will be
performed using Magnetocosmics and CORSIKA (COsmic Ray SImulations for
KAscade)\,\cite{heck_corsika:_1998}. It is a detailed Monte Carlo program to
study the evolution and properties of extensive air showers in the atmosphere.
\textit{Magnetocosmics} \cite{Desorgher2003}, is a code based on Geant4
\cite{Pia2003,Allison2006} that allows the calculation of charged particles
trajectories through different geomagnetic field models.

\subsection{Rigidity of a particle} \label{sec:rigidity_particle}

The motion of a charged particle through a magnetic field is described by the
relativistic Lorentz equation of motion, 
that conserves the magnitude of the momentum $p$, and therefore the energy of
the particle. After some transformations, it can be written as 
$
\frac{d\boldsymbol{\hat{I}_v}}{ds}=\frac{q}{pc}\boldsymbol{\hat{I}_v}\times\boldsymbol{\vec{B}},
$
where $\boldsymbol{\hat{I}_v}$ is a unitary vector pointing in the direction of
the momentum and $s$ is the path length along the particle trajectory. With
this, the rigidity of the particle is defined by:
\begin{equation}
R=\frac{pc}{q},
\end{equation}
and it is a measure of the resistance of the particle to the bending of its
trajectory by the magnetic field.

\subsection{Geomagnetic Rigidity Cut-off}\label{sec:Earth_geomagnetic_models}

The International Geomagnetic Reference Field (IGRF)
\cite{susan_macmillan_international_2010} is an internationally agreed and
widely used mathematical model of the Earth magnetic field up to $5 R_\oplus$.
In this model, the magnetic field vector is given by
$\boldsymbol{\vec{B}}=-\boldsymbol{\vec{\nabla}}V$, where $V$
is the so called magnetic potential. Each constituent model of the IGRF is a set of
spherical harmonics of degree $n$ and order $m$, representing a solution to
Laplace's equation for the magnetic potential arising from sources inside the
Earth at a given epoch. These harmonics are associated with the Gauss
coefficients $g_n^m$ and $h_n^m$, and are updated every five years by the
International Association of Geomagnetism and Aeronomy working group
\cite{susan_macmillan_international_2010}.


Beyond five Earth radii, the Earth magnetic field is increasingly affected by
the solar wind interaction with the Earth magnetosphere. These distortions can
be described by several external source fields originated on different
magnetospheric current systems. The Tsyganenko model\cite{Tsyganenko2002} is a
semi-empirical best-fit representation for the magnetic field, based on a large
number of satellite observations (IMP, HEOS, ISEE, POLAR, Geotail, etc)
\cite{woodfield_comparison_2007}. The model includes the contributions from
external magnetospheric sources such as the ring current, the magnetotail
current system, the magnetopause currents and the large-scale system of
field-aligned currents.

By virtue of the geomagnetic field, it is usual to define the {\bf{rigidity
cut-off}} $R_c$ of a cosmic ray to the lower rigidity of an incoming charged
particle above which it can penetrate the Earth magnetosphere and reach a
specific position at some altitude on the Earth. The rigidity cut-off is
directional, i.e., it depends on the Earth location of the observational point
(characterized by the local altitude $h$, latitude $\varphi$ and longitude
$\lambda$), and the arrival direction of the particle (given by the particle
zenith, $\theta$, and azimuthal, $\phi$, angles): $R_c=R_c(h, \varphi, \lambda,
\theta, \phi)$.

\section{Background Radiation on Board}\label{sec:strategy}

\begin{figure}[t!]
\centering
\includegraphics[scale=.14]{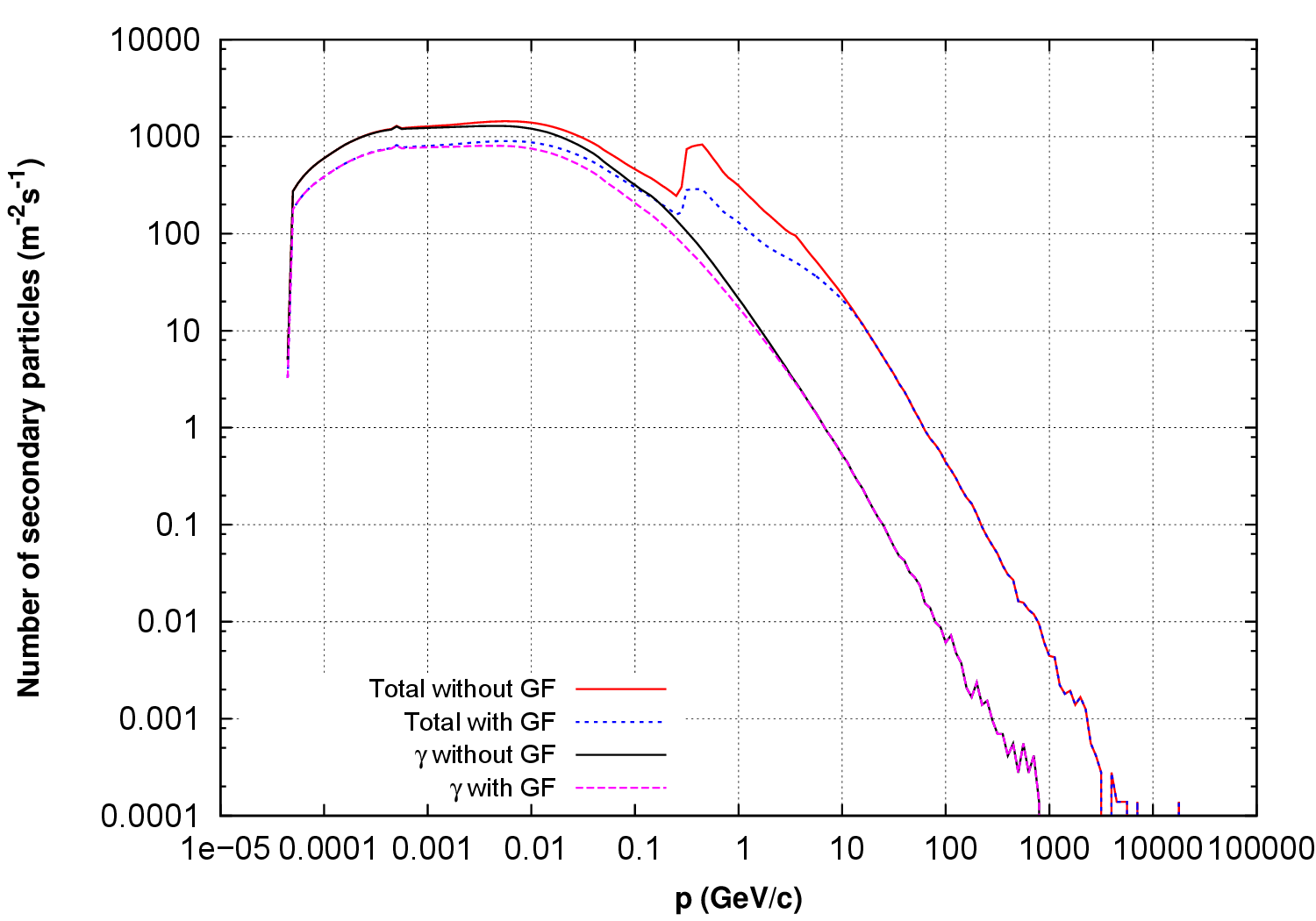}
\caption{Differential spectrum of the secondary particles flux at
$14.74^\circ$S $67.27^\circ$W and an altitude of $11$\,km a.s.l. The effect of
the geomagnetic field on the flux (dashed lines) is clearly visible on the
low energy photon flux, when compared with the flux obtained when no rigidity
cut-off is considered (solid lines).}\label{fig:geo_effect}
\end{figure}

To calculate the expected flux of secondary particles in any place along the
plane trajectory, we use a method based on the simulation of the complete flux
of primaries within a given range of energy \cite{Asorey2011a}, that includes the effect of the rigidity cut-off at different locations in the Earth, that we summarize here: 

\begin{enumerate}
  \item Simulation of showers at different altitudes using \textit{CORSIKA}.
    Features of injected primaries at the top of the atmosphere:
    \begin{itemize}
      \item Primary nuclei injected: $1\leq Z_p\leq 26$, $1\leq A_p\leq 56$
      \item Very low initial rigidity cut-off rigidity: $R_c=4GV$
      \item Energy and arrival direction: $(R_c\times Z_p)\leq (E_p/GeV)\leq
        10^6$, $0^\circ\leq\theta_p\leq 90^\circ$, $0^\circ\leq\phi_p\leq
        360^\circ$
      \item Simulation time: $t=7200 s$ (primary particles flux is constant and
        isotropic)
    \end{itemize}
  \item Selection and discretization of routes.
  \item Computation of rigidity cut-offs for each point in the trajectory using
    \textit{Magnetocosmics}.
  \item Filter secondary particles by the primary rigidities and the rigidity
    cut-off computed for each point of the trajectory: all those showers
    generated by primary particles with rigidities below the cut-off are simply
    discarded.
  \item Computation of the total amount of particles that hit the aircraft, by
    point-to-point integration of the flux of secondaries along the flight
    trajectory.
\end{enumerate}

As an example of the possible results obtained by this method, we show in figure \ref{fig:geo_effect} the differential momentum spectrum of secondary particles
flux obtained when the geomagnetic effect is taken into account, as compared
with the flux when no geomagnetic rigidity cut-off is considered. As expected,
a considerable reduction of the flux of low energy particles is observed, as
the high flux of low energy primaries is strongly diminished and even forbidden
due to the effect of geomagnetic field.

\section{First results}\label{sec:first_results}

\begin{figure}[t!]
\centering
\includegraphics[scale=.14]{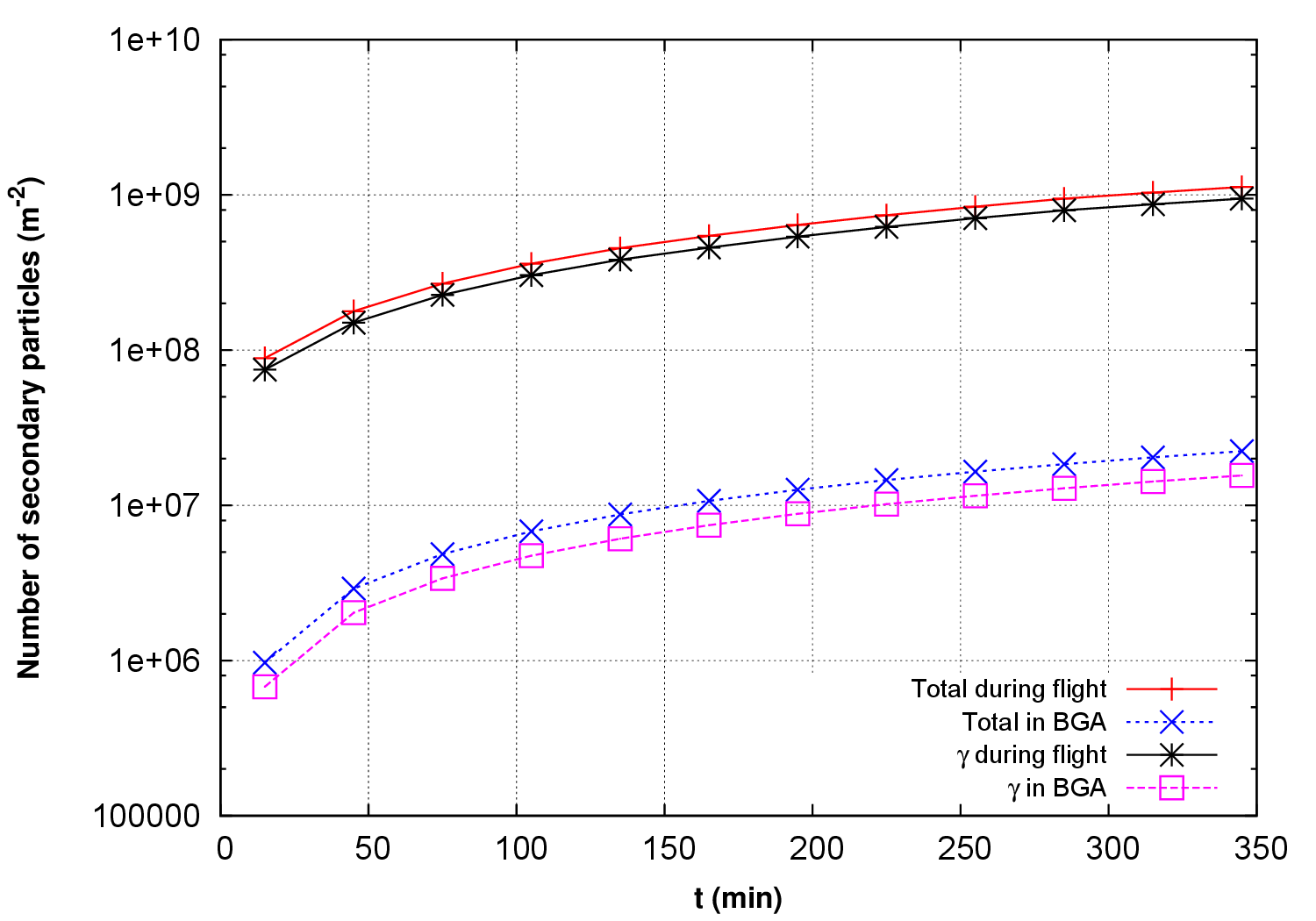}
\caption{Integrated flux of particles (red plus signs) and only photons (black
	asterisks) as a function of time for the flight BOG-EZE, without taken into
	account the fuselage shield and the effects produced by takeoff and
	landing. As a comparison, we show the total integrated flux (blue crosses)
	and photons (magenta squares) for the same calculation but staying the same
	time at the city of Bucaramanga (Colombia), at an altitude of $1$\,km
	a.s.l. There is up to two orders of magnitude in the integrated exposure
	between those two considered cases.}
\label{fig:integrated_bta_bsas}
\end{figure}

In this work we chose the trajectory of the flight AR1360 BOG-EZE
(Bogot\'a-Buenos Aires), and can be seen elsewhere\footnote{See for example
http://www.flightradar24.com/data/flights/ar1360/}. This route was divided into
$12$ intervals of equal flight time and the flux of secondaries along each of
them was assumed to be constant and equal to the flux in the midpoint. The
shield due to the flight fuselage and the effects of takeoff and landing on the
flux was not included in this preliminary analysis (i.e., the aircraft was
supposed to fly at a constant altitude of $11$ km along the whole trajectory).
For each one of this intermediate points, the geomagnetic rigidity cut-off was
calculated by using the method described in the previous section. The result of
this calculation can be seen in figure \ref{fig:integrated_bta_bsas}, where
we show the integrated flux of total particles and photons as a function of
time expected on board of a commercial flight. As a comparison, in the same
figure we show the integrated flux expected for the same calculation but
staying for the same time at the city of Bucaramanga (Colombia) at an altitude
of $1000$\,m a.s.l. In the same figure the effect of the atmospheric absorption
on the EM component is clearly visible as a diminish of the photon flux at
Bucaramanga when compared with the total flux respect to the diminution at
$11$\,km a.s.l..  

\section{Conclusions and acknowledgements}
\label{sec:conclusions}

The simulations performed show that at flight level, the integrated number of
secondary particles is up to two orders of magnitude greater than at $1000$\,m
a.s.l. When calculating the integrated flux, geomagnetic effects must be taken
into account since they reduce the number of primary particles that generate
showers. To make more accurate calculations, the shielding due to aircraft's
fuselage and the effect of takeoff and landing will be included in the
following round of calculations. This calculations also allow us to precisely
calculate the expected flux variations due to space weather phenomena, such as
the observed changes in the flux of primaries during transient geomagnetic
disturbances and the corresponding change in the flux of secondaries. The
authors of this work thank the support of COLCIENCIAS grant 617/2014 and
CDCHT-ULA project C-1598-08-05-A.

\bibliographystyle{elsarticle-num}
\bibliography{biblio_pinilla}

\end{document}